\documentclass{ws-ijmpd}
\usepackage{amssymb}
\usepackage{graphicx}
\begin{document}

%

\def\nocropmarks{\vskip5pt\phantom{cropmarks}}

\let\trimmarks\nocropmarks      

\markboth{R.Aurich and F.Steiner}
{Quintessence and the Curvature of the Universe after WMAP}

%
%

\title{Quintessence and the Curvature of the Universe after WMAP}

\author{Ralf Aurich and Frank Steiner\footnote{
{\it aurich@physik.uni-ulm.de}
and {\it frank.steiner@physik.uni-ulm.de}}}

\address{
Abteilung Theoretische Physik, Universit\"at Ulm,\\
Albert-Einstein-Allee 11,\\ D-89069 Ulm, Germany
}

\maketitle

\pub{Received (received date)}{Revised (revised date)}

\begin{abstract}
We study quintessence models with a constant (effective) equation of state.
It is investigated whether such quintessence models are consistent with
a negative spatial curvature of the Universe with respect to the anisotropy
of the cosmic microwave background radiation measured by the WMAP mission.
If the reionization is negligibly small,
it is found that such models with negative curvature
are admissible due to a geometrical degeneracy.
However, a very high optical depth $\tau$
to the surface of last scattering,
as indicated by the polarization measurements of WMAP,
would rule out such models.
\end{abstract}


\section{Introduction}

During the last decade, observational cosmology has made
enormous progress.
In particular, the spectacular discovery in 1992 of the
temperature fluctuations of the cosmic microwave background (CMB)
radiation by COBE \cite{Smoot_et_al_1992} has provided important clues
about the early Universe and its time evolution.
Although the measured anisotropy of the CMB
(being of the order of $10^{-5}$) is quite small,
it sheds important light on several basic physical mechanisms
(see Refs.\ \cite{Krauss_2003,Kosowsky_2003,Wang_Tegmark_Jain_Zaldarriaga_2002}
for reviews summarizing the last stand before WMAP).
The first findings of NASA's explorer mission
``Wilkinson Microwave Anisotropy Probe'' (WMAP)
\cite{Bennett_et_al_2003}
has tremendously increased our knowledge of the temperature fluctuations
of the CMB, since WMAP has measured the anisotropy of the CMB radiation
over the full sky with high accuracy.
It is the purpose of this paper to test our theoretical
understanding of physical cosmology by making use of the
high quality of the WMAP data.

One of the most important results of the cosmological observations made
during the last few years has been that all tests of cosmology
probing the nature of matter resp.\ energy in the Universe
suggest that our Universe is at the present epoch not dominated by
matter but rather by a mysterious energy with negative pressure
(see Ref.\ \cite{Krauss_2003} for a recent review
of the state of the Universe).
There are essentially two models to explain this new energy component
which nowadays is called dark energy.
One possibility is to identify the dark energy with Einstein's
cosmological constant $\Lambda$ with a corresponding energy density
$\varepsilon_\Lambda=\Lambda c^4/(8\pi G)$ and
negative pressure $p_\Lambda=-\varepsilon_\Lambda$,
assuming a positive cosmological constant.
An alternative explanation is quintessence
\cite{Turner_1983,Peebles_Ratra_1988,Ratra_Peebles_1988,%
Caldwell_Dave_Steinhardt_1998,Aurich_Steiner_2002a},
where the dark energy is identified with the energy density
$\varepsilon_\phi$ arising from a time-evolving scalar
(quintessence) field $\phi$
(see Refs.\ \cite{Sahni_Starobinsky_2000,Peebles_Ratra_2002}
for recent reviews).
An important cosmological parameter is the equation of state of
the dark energy component,
which in the case of quintessence is given by
$w_\phi := p_\phi/\varepsilon_\phi$,
where $p_\phi$ denotes the associated pressure.
Quintessence models can be considered as generalizations
of the cosmological constant and include the latter as the special case
$w_\phi = \hbox{const.} = -1$.
In general, the equation of state $w_\phi$ can be a function of redshift.
In our following analysis of the WMAP data,
we shall assume, however, $w_\phi = \hbox{const.} \geq -1$.
These models have the property that their quintessence component
is negligible at early times such as the recombination epoch.
(There are also models with a time-dependent $w_\phi$
which lead to ``early quintessence'' models
\cite{Caldwell_Doran_Mueller_Schaefer_Wetterich_2003},
which are, however, not discussed in this paper.)
First of all, a constant equation of state for the dark energy
component represents an obvious and also the simplest generalization
of the other known energy components of the Universe
which have the constant values $w_{\hbox{\scriptsize r}}=1/3$ (for radiation)
and $w_{\hbox{\scriptsize m}}=0$ (for matter).
Secondly, as discussed e.\,g.\ in Refs.\
\cite{Maor_Brustein_Steinhardt_2001,Doran_Lilley_Schwindt_Wetterich_2001,%
Cornish_2000,Huterer_Turner_2001,Aurich_Steiner_2002b},
there is an inherent theoretical limitation to determine the time variation
of the equation of state of the dark energy component.
For example, the luminosity distance $d_L$ and the
angular-diameter distance $d_A$ depend on $w_\phi$
through a multiple-integral relation that smears out detailed
information about the redshift dependence of $w_\phi(z)$
(see Eqs.\,(\ref{Eq:d_A}) and (\ref{Eq:Hubble}) below).

One of the most fundamental cosmological parameters is the curvature
of the Universe.
Many cosmologists seem to accept as established that the Universe
is flat corresponding to $k=0$ and
$\Omega_{\hbox{\scriptsize tot}} :=
\varepsilon_{\hbox{\scriptsize tot}} /
\varepsilon_{\hbox{\scriptsize crit}}=1$.
Contrary to that opinion, we shall demonstrate
that the WMAP data are {\it consistent}
with certain quintessence models possessing a constant equation of state
in a hyperbolic universe, i.\,e.\ with {\it negative} spatial curvature,
$k=-1$, corresponding to $\Omega_{\hbox{\scriptsize tot}} < 1$,
provided that the optical depth $\tau$ is negligibly small.
Our result is the consequence of the crucial observation that
there exists a {\it degeneracy} in the space of the relevant cosmological
parameters $(\Omega_{\hbox{\scriptsize tot}}, \Omega_\phi, w_\phi)$,
where $\Omega_\phi := \varepsilon_\phi /
\varepsilon_{\hbox{\scriptsize crit}}$ is the ratio of the quintessence
energy density to the critical density.

\section{Quintessence models}

Our background model is the standard cosmological model based on a
Friedmann--Lema\^{\i}tre universe with Robertson-Walker metric.
Then the Friedmann equation reads $(a' := da/d\eta, c=1)$
\begin{equation}
\label{Eq:Friedmann}
H^2 := \left(\frac{a'}{a^2}\right)^2 \; = \;
\frac{8\pi G}3 \varepsilon_{\hbox{\scriptsize tot}} - \frac k{a^2}
\hspace{10pt} ,
\end{equation}
where $a(\eta)$ is the cosmic scale factor as a function of
conformal time $\eta$ and $H=H(\eta)$ is the Hubble parameter.
The last term in Eq.~(\ref{Eq:Friedmann}) is the curvature term.
Furthermore, $\varepsilon_{\hbox{\scriptsize tot}} :=
\varepsilon_{\hbox{\scriptsize r}} +
\varepsilon_{\hbox{\scriptsize m}} + \varepsilon_\phi$,
where $\varepsilon_{\hbox{\scriptsize r}}$ denotes the energy density of
``radiation'', i.\,e.\ of the relativistic components according to
photons and three massless neutrinos;
$\varepsilon_{\hbox{\scriptsize m}} =
\varepsilon_{\hbox{\scriptsize b}} +
\varepsilon_{\hbox{\scriptsize cdm}}$
is the energy density of non-relativistic ``matter'' consisting of
baryonic matter, $\varepsilon_{\hbox{\scriptsize b}}$, and
cold dark matter, $\varepsilon_{\hbox{\scriptsize cdm}}$,
and $\varepsilon_\phi$ is the energy density of the dark energy
due to the quintessence field $\phi$.

In quintessence models, the energy density $\varepsilon_\phi(\eta)$
and the pressure $p_\phi(\eta)$ of the dark energy are determined by
the quintessence potential $V(\phi)$
\begin{equation}
\label{Eq:eos_phi}
\varepsilon_\phi \; = \; \frac{1}{2 a^2} \, {\phi'}^2 + V(\phi)
\hspace{10pt} , \hspace{10pt}
p_\phi \; = \; \frac{1}{2 a^2} \, {\phi'}^2 - V(\phi)
\hspace{10pt} ,
\end{equation}
or equivalently by the equation of state
$w_\phi = \frac{p_\phi}{\varepsilon_\phi}$.
The equation of motion of the real, scalar field $\phi(\eta)$ is
\begin{equation}
\label{Eq:equation_of_motion_phi}
\phi'' \, + \, 2 \frac{a'}{a} \phi' \, + \,
a^2 \frac{\partial V(\phi)}{\partial \phi} \; = \; 0
\hspace{10pt} ,
\end{equation}
where it is assumed that $\phi$ couples to matter only through gravitation.
The various energy densities are constrained by the continuity equation
\begin{equation}
\label{Eq:continuity_equation}
\varepsilon_{\hbox{\scriptsize x}}' \, + \,
3 (1+w_{\hbox{\scriptsize x}}) \, \frac{a'}a \,
\varepsilon_{\hbox{\scriptsize x}} \; = \; 0
\hspace{10pt} ,
\end{equation}
with the constant equation of state
$w_{\hbox{\scriptsize r}}=\frac 13$,
$w_{\hbox{\scriptsize m}}=0$ for $\hbox{x}=\hbox{r},\hbox{m}$,
respectively, and $w_\phi(\eta)$ for $\hbox{x}=\phi$.
It is worthwhile to remark that the quintessence field $\phi$ may be regarded
as a real physical field, or simply as a device for modeling more general
cosmic fluids with negative pressure.
Since $w_\phi$ is assumed to be constant, the potential $V(\phi)$
is uniquely determined as shown in
Refs.\ \cite{Aurich_Steiner_2002a,Aurich_Steiner_2002b}.
However, in solving numerically the coupled
Eqs.\,(\ref{Eq:Friedmann})-(\ref{Eq:continuity_equation})
no knowledge of the potential $V(\phi)$ is required.

\section{Comparison with the WMAP data}

In our comparison with the WMAP data, we make the following assumptions
\cite{Krauss_2003}.
The Hubble constant $H_0 = h \times 100 \hbox{ km s}^{-1} \hbox{Mpc}^{-1}$
is set to $h=0.70$, and the baryonic density parameter
$\Omega_{\hbox{\scriptsize b}}=0.05$
(i.\,e.\ $\Omega_{\hbox{\scriptsize b}} h^2 = 0.0245$)
is chosen in agreement with the current Big-Bang nucleosynthesis constraints.
Furthermore, the initial curvature perturbation is assumed to be
scale-invariant $(n_S=1)$ which is suggested by inflationary models.
We have also studied models having a spectral index $n_S=0.95$,
but the WMAP data alone,
i.\,e.\ not taking into account large-scale structure data,
prefer models having $n_S=1$.
(We do not consider a $k$-dependent spectral index $n_S$ in this paper.)
In addition, we have varied the Hubble constant over the
range $h=0.64\dots 0.72$, but the best models for a fixed curvature
all prefer a value close to $h=0.70$.
The optical depth $\tau$ to the surface of last scattering is
in this Section assumed to be negligibly small.
In the next Section, we use the surprisingly high value of $\tau$
found by WMAP\cite{Kogut_et_al_2003}.
The CMB anisotropy of the quintessence models is computed using
the publicly available CAMB code by Antony Lewis and Antony Challinor
together with the quintessence module for $w_\phi=\hbox{const.}$ models.
The angular power spectrum $\delta T_l^2 = l(l+1)C_l/2\pi$
is compared with the one obtained by WMAP \cite{Hinshaw_et_al_2003}
which comprises 899 data points.
The amplitude of the initial curvature perturbation is determined
such that the value of $\chi_{\hbox{\scriptsize eff}}^2$ is minimized.
The values of $\chi_{\hbox{\scriptsize eff}}^2$ are obtained by
using the FORTRAN code (provided by the WMAP team \cite{Verde_et_al_2003})
which, given a theoretical CMB power spectrum,
computes the likelihood of that model fit to the WMAP data.

\begin{figure}[htb]
\begin{center}
\begin{minipage}{8cm}
\includegraphics[width=8.0cm]{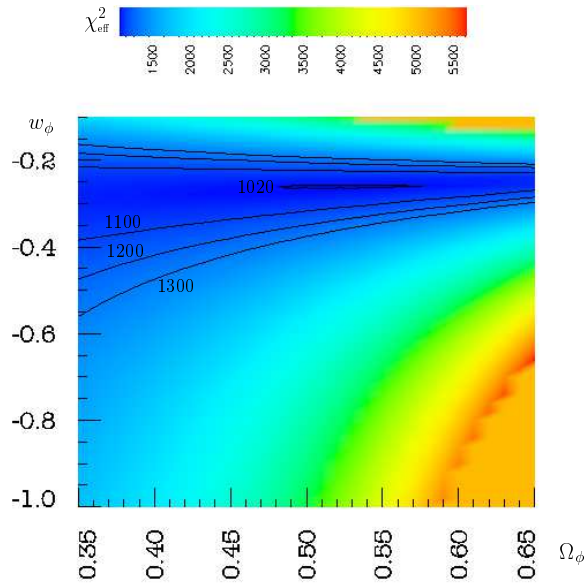}
\end{minipage}
\vspace*{-8pt}
\end{center}
\caption{\label{Fig:Omega_tot_85_2d}
The $\chi^2$ values are shown in dependence on $w_\phi$ and $\Omega_\phi$
for $\Omega_{\hbox{\scriptsize tot}} = 0.85$,
$h=0.70$, $\tau=0.0$ and $n_s = 1.0$
using the WMAP data.
The curves with $\chi_{\hbox{\scriptsize eff}}^2=1020$, 1100, 1200
and 1300 are indicated.
The minimum $\chi^2_{\hbox{\scriptsize min}} = 1016$ occurs at
$\Omega_\phi=0.54$, $w_\phi=-0.26$.
}
\end{figure}

\begin{figure}[htb]
\begin{center}
\begin{minipage}{8cm}
\includegraphics[width=8.0cm]{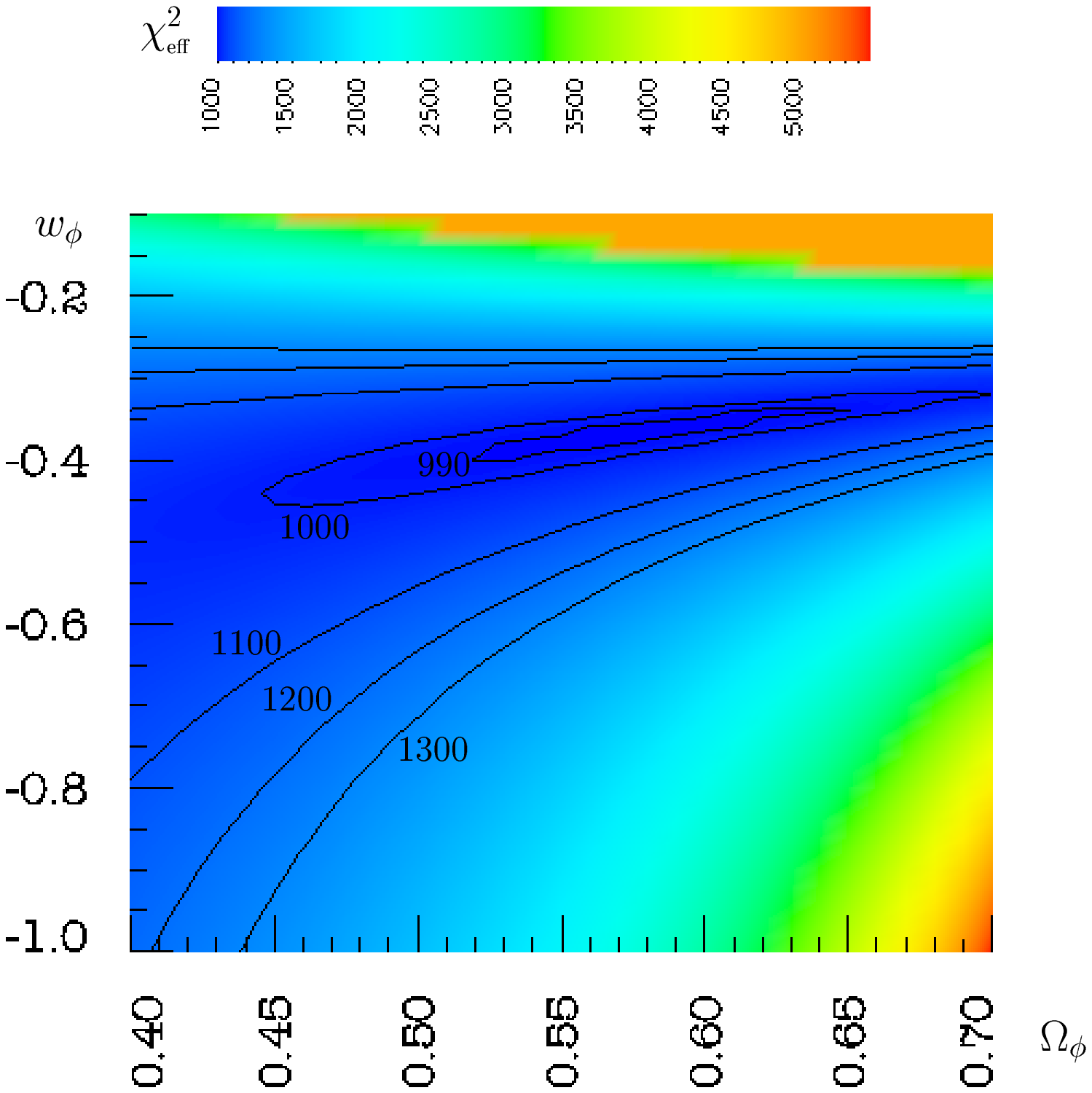}
\end{minipage}
\vspace*{-8pt}
\end{center}
\caption{\label{Fig:Omega_tot_90_2d}
The same as in Fig.\,\ref{Fig:Omega_tot_85_2d}
for $\Omega_{\hbox{\scriptsize tot}} = 0.9$.
The curves with $\chi_{\hbox{\scriptsize eff}}^2=990$, 1000, 1100, 1200
and 1300 are indicated.
The minimum $\chi^2_{\hbox{\scriptsize min}} = 985$ occurs at
$\Omega_\phi=0.59$, $w_\phi=-0.36$.
}
\end{figure}
\begin{figure}[htb]
\begin{center}
\begin{minipage}{8cm}
\includegraphics[width=8.0cm]{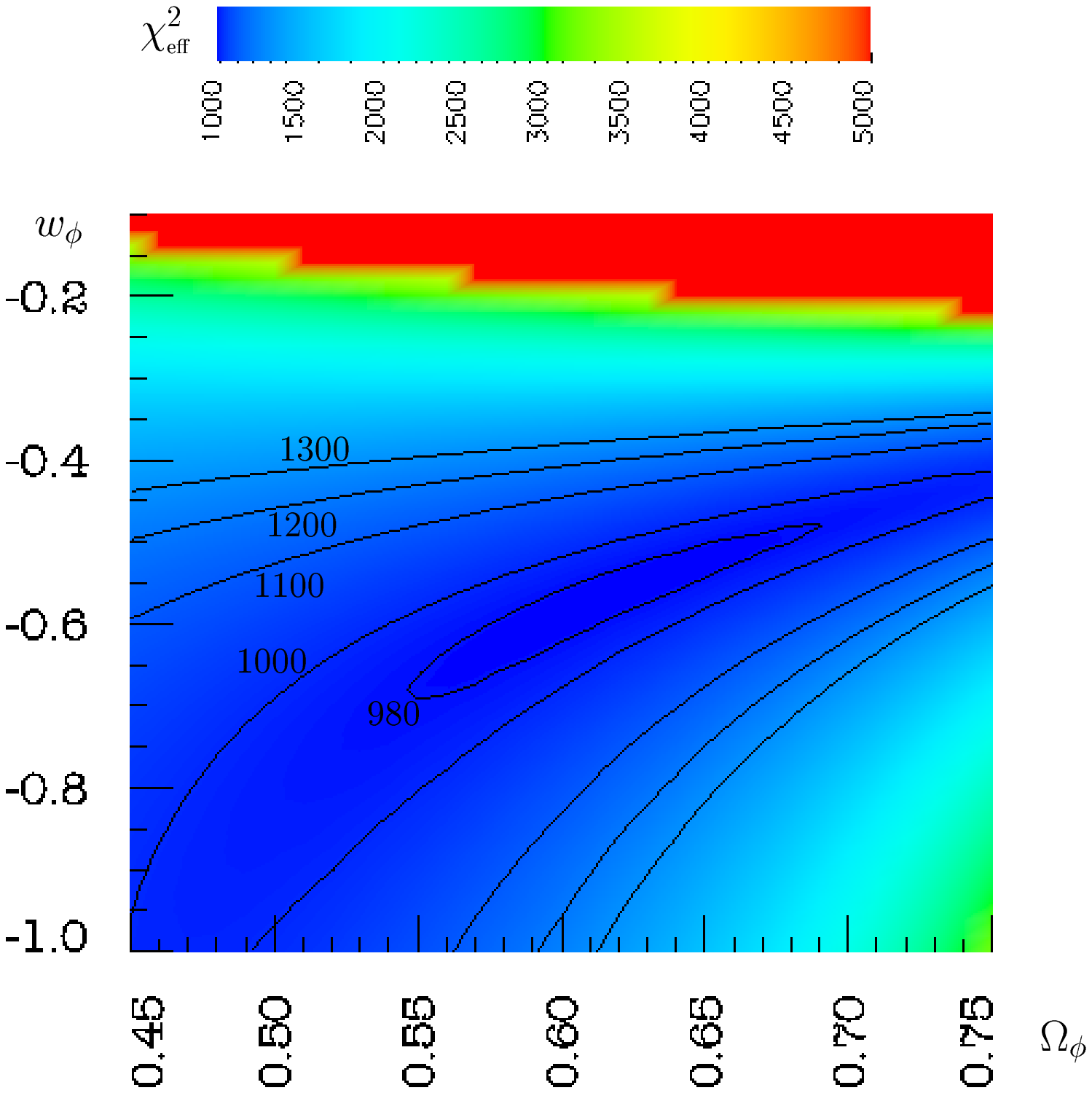}
\end{minipage}
\vspace*{-8pt}
\end{center}
\caption{\label{Fig:Omega_tot_95_2d}
The same as in Fig.\,\ref{Fig:Omega_tot_85_2d}
for $\Omega_{\hbox{\scriptsize tot}} = 0.95$.
The curves with $\chi_{\hbox{\scriptsize eff}}^2=980$, 1000, 1100, 1200
and 1300 are indicated.
The minimum $\chi^2_{\hbox{\scriptsize min}} = 975$ occurs at
$\Omega_\phi=0.61$, $w_\phi=-0.58$.
}
\end{figure}
\begin{figure}[htb]
\begin{center}
\begin{minipage}{8cm}
\includegraphics[width=8.0cm]{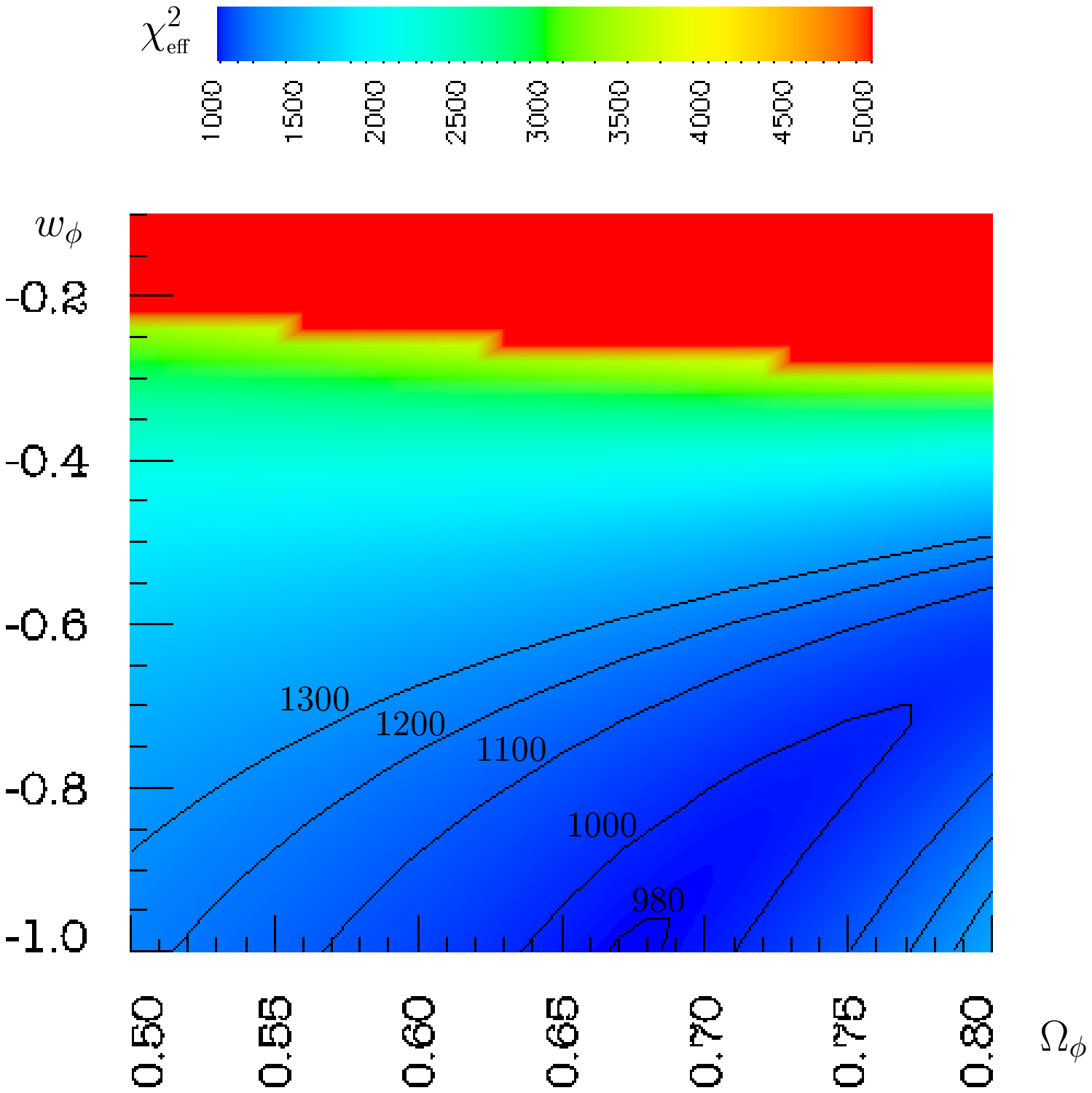}
\end{minipage}
\vspace*{-8pt}
\end{center}
\caption{\label{Fig:Omega_tot_1_2d}
The same as in Fig.\,\ref{Fig:Omega_tot_85_2d}
for $\Omega_{\hbox{\scriptsize tot}} = 1.0$.
The curves with $\chi_{\hbox{\scriptsize eff}}^2=980$, 1000, 1100, 1200
and 1300 are indicated.
The minimum $\chi^2_{\hbox{\scriptsize min}} = 978$ occurs at
$\Omega_\phi=0.68$, $w_\phi=-1.0$.
}
\end{figure}

\begin{figure}[htb]
\begin{center}
\begin{minipage}{10cm}
\includegraphics[width=10.0cm]{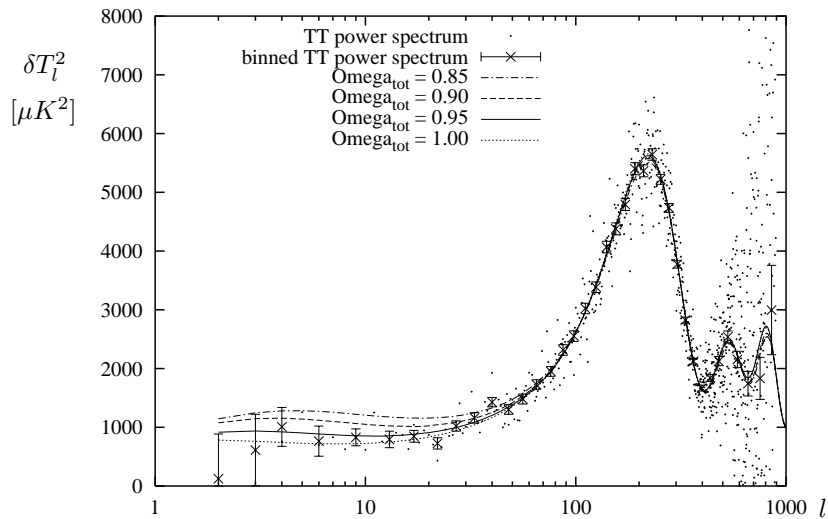}
\put(-1,4){$l$}
\put(-302,172){$\delta T_l^2$}
\put(-307,155){$[\mu K^2]$}
\end{minipage}
\vspace*{-10pt}
\end{center}
\caption{\label{Fig:aps_best_w_const_log}
The angular power spectrum $\delta T_l^2$ is presented for the
four best models for the cases
$\Omega_{\hbox{\scriptsize tot}} = 0.85, 0.9, 0.95$ and
$\Omega_{\hbox{\scriptsize tot}} = 1.0$ and compared
with the WMAP\protect\cite{Bennett_et_al_2003} data.
The model fits are determined by the minima of
$\chi_{\hbox{\scriptsize eff}}^2$
shown in Figures \ref{Fig:Omega_tot_85_2d} to \ref{Fig:Omega_tot_1_2d}.
}
\end{figure}

\begin{figure}[htb]
\begin{center}
\begin{minipage}{10cm}
\includegraphics[width=10.0cm]{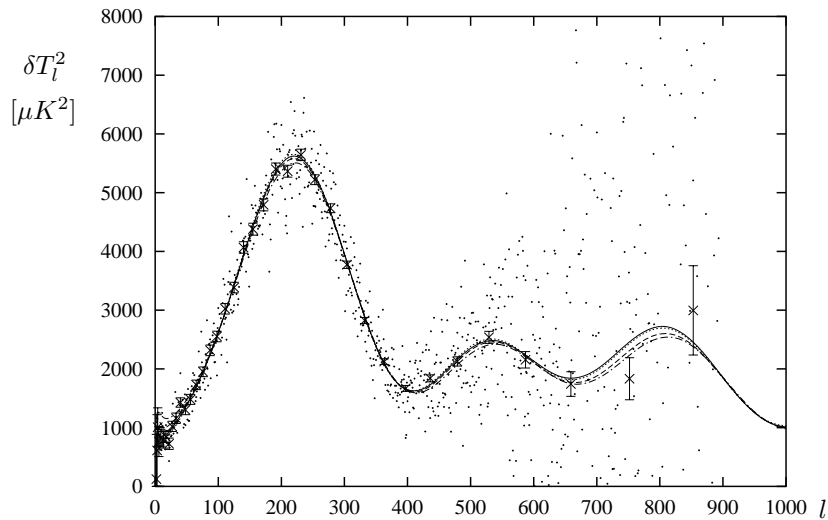}
\put(-1,4){$l$}
\put(-302,172){$\delta T_l^2$}
\put(-307,155){$[\mu K^2]$}
\end{minipage}
\vspace*{-10pt}
\end{center}
\caption{\label{Fig:aps_best_w_const_lin}
The same angular power spectra $\delta T_l^2$ as in
Figure \ref{Fig:aps_best_w_const_log} are presented in a
linear $l$-scale.
}
\end{figure}

Figures \ref{Fig:Omega_tot_85_2d} to \ref{Fig:Omega_tot_1_2d} present
the values of $\chi_{\hbox{\scriptsize eff}}^2$ for the four cases
$\Omega_{\hbox{\scriptsize tot}} = 0.85, 0.9, 0.95$ and $1.0$, respectively,
in dependence on $w_\phi$ and $\Omega_\phi$.
Since $\Omega_{\hbox{\scriptsize tot}}$ is held fixed,
the matter density $\Omega_{\hbox{\scriptsize m}} :=
\Omega_{\hbox{\scriptsize b}} + \Omega_{\hbox{\scriptsize cdm}}$
is given by
$\Omega_{\hbox{\scriptsize m}} = \Omega_{\hbox{\scriptsize tot}} -
\Omega_\phi$ neglecting the small radiation contribution.
One observes that a decreasing $\Omega_{\hbox{\scriptsize tot}}$
demands an increasing $w_\phi$, i.\,e.\ a less negative value;
from $\Omega_{\hbox{\scriptsize tot}} = 0.85$ to
$\Omega_{\hbox{\scriptsize tot}} = 1.0$ the equation of state
changes from $w_\phi=-0.26$ to $w_\phi=-1.0$.
Among the models considered, the best fit to the WMAP data
is obtained for the model with
$\Omega_{\hbox{\scriptsize tot}} = 0.95$,
$\Omega_\phi=0.61$ and $w_\phi=-0.58$ belonging to the minimum at
$\chi_{\hbox{\scriptsize eff}}^2=975$,
see Figure \ref{Fig:Omega_tot_95_2d}.
The corresponding angular power spectrum
$\delta T_l^2$ is shown as a full curve in Figures
\ref{Fig:aps_best_w_const_log} and \ref{Fig:aps_best_w_const_lin}
and compared with the WMAP measurements.
The model with $\Omega_{\hbox{\scriptsize tot}} = 0.90$,
$\Omega_\phi=0.59$ and $w_\phi=-0.36$ is almost of the same quality
and belongs to the minimum at $\chi_{\hbox{\scriptsize eff}}^2=985$
(see Figure \ref{Fig:Omega_tot_90_2d})
compared to the 899 data points of WMAP.
The corresponding $\delta T_l^2$ is shown as dashed curve in Figures
\ref{Fig:aps_best_w_const_log} and \ref{Fig:aps_best_w_const_lin}.
Assuming a flat universe, $\Omega_{\hbox{\scriptsize tot}} = 1.0$,
one obtains a broad, but extremely flat
$\chi_{\hbox{\scriptsize eff}}^2$-valley
as shown in Figure \ref{Fig:Omega_tot_1_2d}.
The best flat model is obtained for
$\Omega_\phi=0.68$ and $w_\phi=-1.0$ and is plotted as dotted curve
in Figures
\ref{Fig:aps_best_w_const_log} and \ref{Fig:aps_best_w_const_lin}.
This model possesses a $\chi_{\hbox{\scriptsize eff}}^2$-value of
$\chi_{\hbox{\scriptsize eff}}^2=978$.
It is worth noting that the $\chi_{\hbox{\scriptsize eff}}^2$-valleys
get increasingly narrower with a horizontal alignment in Figures
\ref{Fig:Omega_tot_85_2d}-\ref{Fig:Omega_tot_1_2d} with decreasing
$\Omega_{\hbox{\scriptsize tot}}$ and thus allow only a small variation
in $w_\phi$.
In the case $\Omega_{\hbox{\scriptsize tot}} = 0.85$,
one observes the most confined valley,
see Figure \ref{Fig:Omega_tot_85_2d},
where the best model with $\Omega_\phi=0.54$ and $w_\phi=-0.26$
has $\chi_{\hbox{\scriptsize eff}}^2=1016$.
It is seen in Figures
\ref{Fig:aps_best_w_const_log} and \ref{Fig:aps_best_w_const_lin}
that the best models corresponding to the cases
$\Omega_{\hbox{\scriptsize tot}} = 0.85, 0.9, 0.95$ and $1.0$, respectively,
are nearly degenerate with respect to their angular power spectrum
$\delta T_l^2$.
The main difference occurs for small values of $l$
where the integrated Sachs-Wolfe effect yields with decreasing
$\Omega_{\hbox{\scriptsize tot}}$ and increasing $w_\phi$
a larger contribution.
However, this discrepancy scarcely influences the value of
$\chi_{\hbox{\scriptsize eff}}^2$ due to the cosmic variance.
In contrast, the position of the acoustic peaks and their relative heights
are nearly indistinguishable.
Even with the high accuracy of the WMAP data,
one can hardly discriminate between these models.

\begin{figure}[htb]
\begin{center}
\begin{minipage}{8cm}
\includegraphics[width=8.0cm]{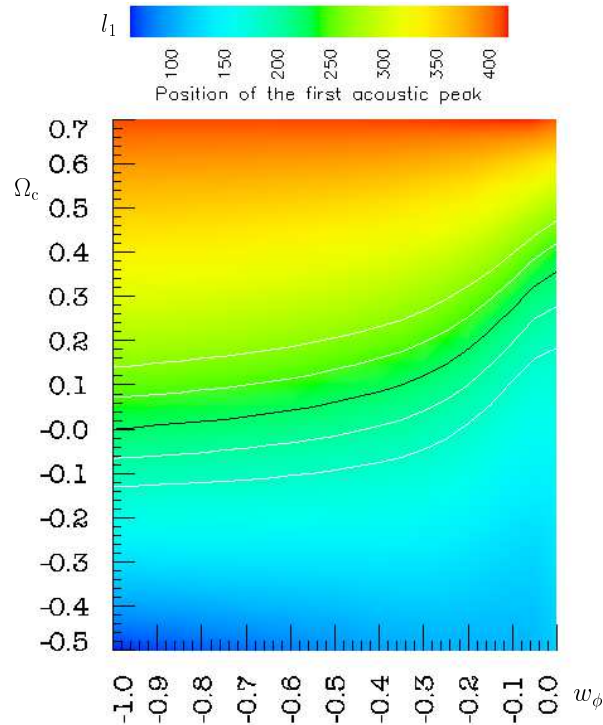}
\end{minipage}
\vspace*{-8pt}
\end{center}
\caption{\label{Fig:Peak_Postion}
The position $l_1$ of the first acoustic peak is computed
in dependence on $w_\phi$ and
$\Omega_{\hbox{\scriptsize c}} = 1 - \Omega_{\hbox{\scriptsize tot}}$
for $\Omega_{\hbox{\scriptsize b}} = 0.05$ and
$\Omega_{\hbox{\scriptsize cdm}} = 0.27$.
The white curves indicate the peak position for $l_1=180, 200, 240$ and 260.
The black curve belongs to $l_1=220$.
}
\end{figure}

The identical position of the peaks
is mainly due to a geometrical degeneracy
\cite{Efstathiou_Bond_1999,Aurich_Steiner_2002b}
which arises through the angular-diameter distance
$d_A$ to the surface of last scattering which has to be computed by
$(\Omega_{\hbox{\scriptsize c}} := 1 - \Omega_{\hbox{\scriptsize tot}})$
\begin{equation}
\label{Eq:d_A}
d_A \; = \;
\frac 1{H_0\sqrt{|\Omega_{\hbox{\scriptsize c}}|}
(1+z_{\hbox{\scriptsize sls}})} \,
S_k\left( H_0\sqrt{|\Omega_{\hbox{\scriptsize c}}|}
\int_1^{1+z_{\hbox{\scriptsize sls}}} \frac{dx}{H(x)} \, \right)
\hspace{5pt} ,
\end{equation}
where $S_k(y)$ denotes $\sinh y$, $y$,  $\sin y$ for $k=-1,0,+1$,
respectively.
The Hubble parameter $H(x)$, $x=z+1$, is given by
\begin{equation}
\label{Eq:Hubble}
\left(\frac{H(x)}{H_0}\right)^2 =  \Omega_{\hbox{\scriptsize r}} x^4 +
\Omega_{\hbox{\scriptsize m}} x^3 -
k\Omega_{\hbox{\scriptsize c}} x^2 +
\Omega_\phi e^{3\int_1^x (1+w_\phi(x')) \frac{dx'}{x'}}
\hspace{2pt} .
\end{equation}
The degeneracy due to $d_A$ has been discussed by several authors
\cite{Efstathiou_Bond_1999,Hu_Eisenstein_Tegmark_White_1999,%
Cornish_2000,Huterer_Turner_2001,Aurich_Steiner_2002b,Bean_Melchiorri_2002,%
Melchiorri_Mersini_Odman_Trodden_2002},
in particular in the neighborhood of a flat universe and for
$w_\phi$ near $-1$.
One emphasis has been on the dependence of the position $l_1$ of the first
acoustic peak in the angular power spectrum of the CMB anisotropy
on the various cosmological parameters.
As a result, it has been concluded that, although there is a degeneracy
with respect to $\Omega_\phi$ and $w_\phi$,
the position of the first peak is well suited for a determination of the
curvature of the Universe.
Analyzing the CMB anisotropy and other cosmological observations
has led many cosmologists to accept
that the Universe is flat in accordance with the inflationary scenarios.
However, admitting also quintessence models with $w_\phi>-1$ changes this
simple picture, since there exists now a degeneracy in the
$(\Omega_{\hbox{\scriptsize tot}}, \Omega_\phi, w_\phi)$-space.
In Ref.\ \cite{Aurich_Steiner_2002b} the dependence of $d_A$ on the
curvature and on $w_\phi$ has been investigated
using the CMB data before WMAP and it has been shown
that there exists a family of models with different curvature
having all the same $d_A$.
In Figure \ref{Fig:Peak_Postion} we show the position $l_1$
of the first acoustic peak (computed using CAMB) for a large
class of models sharing the cosmological parameters
$\Omega_{\hbox{\scriptsize cdm}} = 0.27$,
$h=0.70$, $\tau=0.0$ and $n_S=1.0$.
The WMAP data confines the position of the first peak close
to $l_1 \simeq 220$ \cite{Hinshaw_et_al_2003}.
The models possessing $l_1 = 220$ are connected by the black curve in
Figure \ref{Fig:Peak_Postion}.
It is worthwhile to note that the curve for $l_1 = 220$
allows flat models as well as models with negative curvature,
but not with positive curvature.
Using the WMAP data, the best models,
shown in Figures \ref{Fig:aps_best_w_const_log} and
\ref{Fig:aps_best_w_const_lin},
belong to models lying close to this black curve.
The existence of this family in the parameter space
$(\Omega_{\hbox{\scriptsize tot}}, \Omega_\phi, w_\phi)$
of equal values of $l_1$ is mainly due to the degeneracy in $d_A$.
Under the assumption that the reionization has no significant
influence on the CMB anisotropy,
we conclude that this geometrical degeneracy allows our Universe to
possess negative spatial curvature compatible with the new WMAP data.

\section{Polarization and the influence of early reionization}

\begin{figure}[b]
\begin{center}
\begin{minipage}{12cm}
\includegraphics[width=6.0cm]{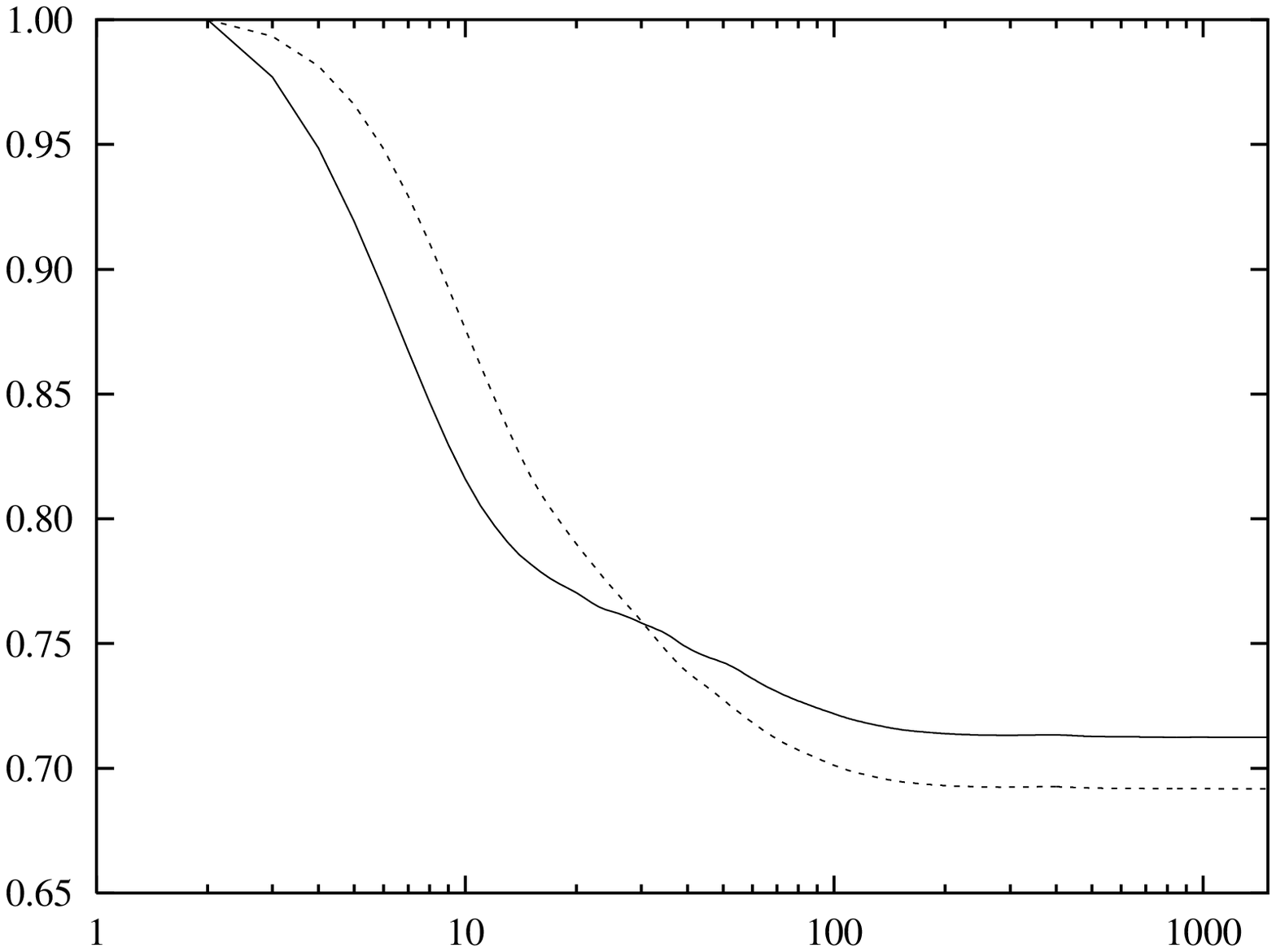}
\put(-30,100){a)}
\put(-172,108){$_{\cal R}$}
\put(-5,4){$_l$}
\hspace*{10pt}\includegraphics[width=6.0cm]{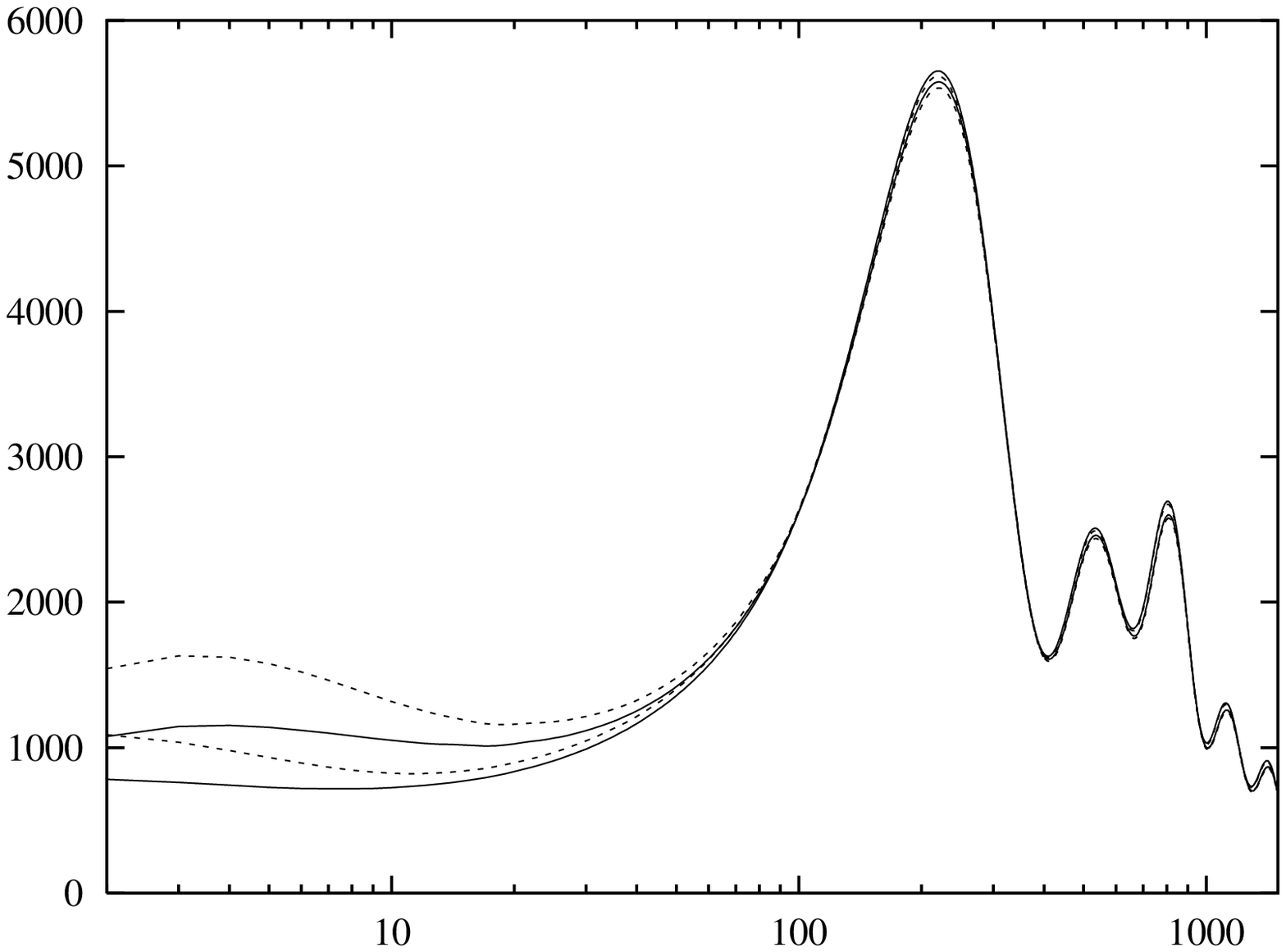}
\put(-135,100){b)}
\put(-5,4){$_l$}
\put(-174,108){$_{\delta T_l^2}$}
\put(-177, 90){$_{[\mu K^2]}$}
\end{minipage}
\vspace*{-10pt}
\end{center}
\caption{\label{Fig:reionization_suppression}
Panel a) shows the reionization suppression
${\cal R} := C_l(\tau=0.15) / C_l(\tau=0.0)$
for the two ``best'' models for
$\Omega_{\hbox{\scriptsize tot}} = 1.0$ (solid curve) and
$\Omega_{\hbox{\scriptsize tot}} = 0.9$ (dashed curve),
if the optical depth $\tau$ is increased from
$\tau =0.0$ to $\tau=0.15$.
Panel b) shows $\delta T_l^2$ for these two cases
with $\tau =0.0$ (solid curves) and $\tau =0.15$ (dashed curves).
The two upper curves at low values of $l$ belong to the case with
$\Omega_{\hbox{\scriptsize tot}} = 0.9$.
}
\end{figure}

WMAP has also obtained data for the
correlation of the polarization with the temperature variations
\cite{Kogut_et_al_2003}.
This data requires an unexpected high optical depth $\tau$
to the surface of last scattering of
$\tau=0.17 \pm 0.04$ \cite{Kogut_et_al_2003} and
$\tau=0.17 \pm 0.07$ \cite{Spergel_et_al_2003}.
This early reionization erases some of the primary anisotropy
below the corresponding horizon (reionization damping) \cite{Hu_White_1997}
which roughly corresponds to $l\simeq 10$ for the considered models.
Thus the angular power spectrum is only slightly altered below $l\simeq 10$
as can be seen in Figure \ref{Fig:reionization_suppression}a)
in the case of our two best models for
$\Omega_{\hbox{\scriptsize tot}} = 1.0$ and
$\Omega_{\hbox{\scriptsize tot}} = 0.9$, respectively.
For $l\gtrsim 100$ the reionization damping is almost independent
of $l$ and, thus, does not alter the peak structure.
The erased power for $l\gtrsim 10$, i.\,e.\, the reduced peak height
(proportional to $e^{-2\tau}$),
is compensated by increasing the initial perturbation amplitude
in order to obtain a first peak of the right amplitude.
As a consequence the multipoles below $l\simeq 10$ are increased in
comparison with models possessing negligible reionization
as shown in Figure \ref{Fig:reionization_suppression}b).
A close inspection of Figure \ref{Fig:aps_best_w_const_log} reveals
that our models with negative curvature
already possess a bit too much power for small values of $l$.
These models have been computed with negligible optical depth $\tau=0$ 
to the surface of last scattering.
Thus increasing the optical depth $\tau$ increases the power
for small values of $l$ so severely that the models
get such high values of $\chi_{\hbox{\scriptsize eff}}^2$
that they are very unlikely.

\begin{figure}[htb]
\begin{center}
\begin{minipage}{10cm}
\includegraphics[width=10.0cm]{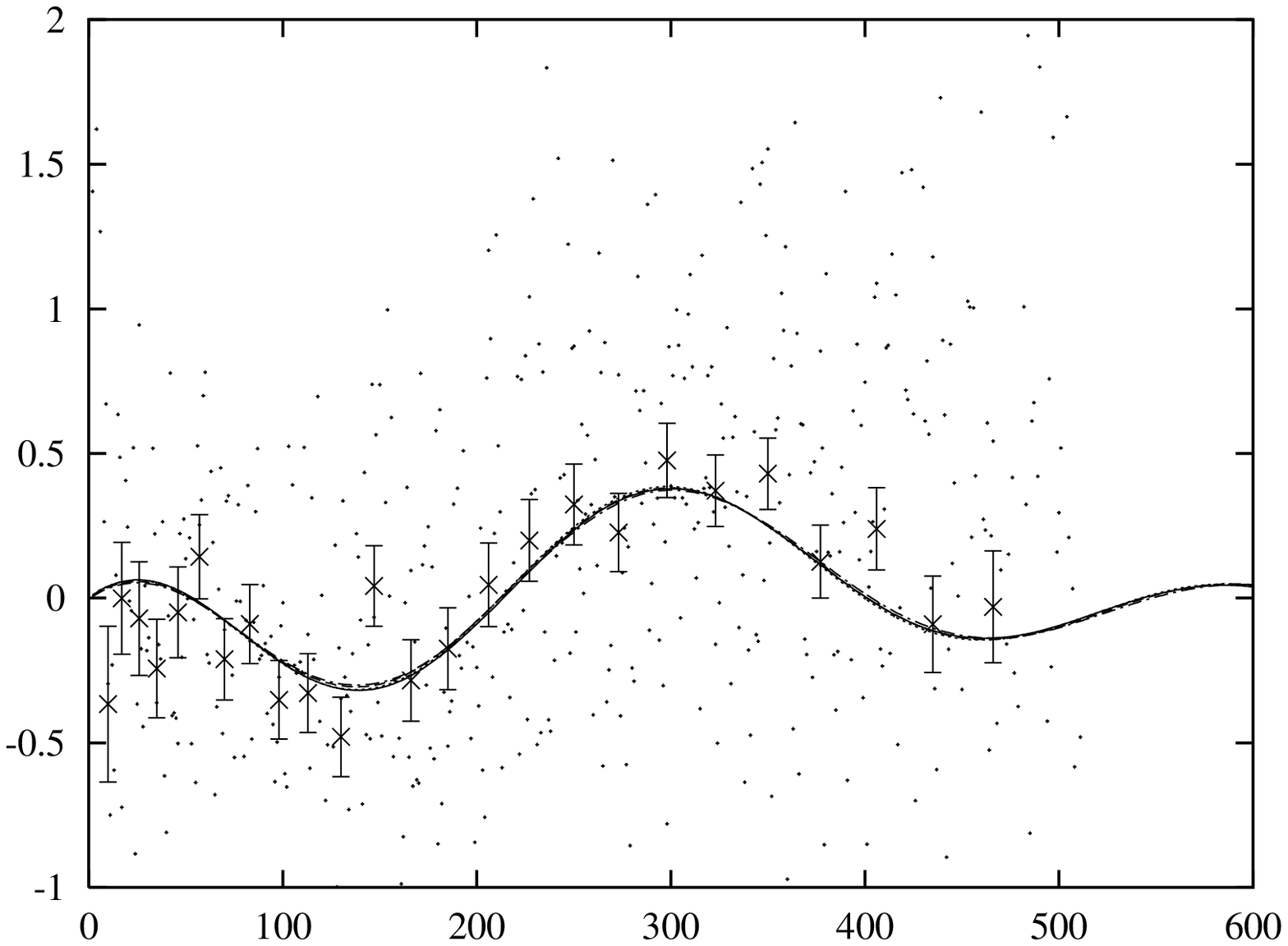}
\put(-3,4){$l$}
\put(-312,182){$\frac{l+1}{2\pi} C_l^{\hbox{\scriptsize{TE}}}$}
\put(-307,165){$[\mu K^2]$}
\end{minipage}
\vspace*{-10pt}
\end{center}
\caption{\label{Fig:aps_best_w_const_lin_pol}
The cross power spectra $(l+1) C_l^{\hbox{\scriptsize{TE}}}/(2\pi)$
are presented for the four best models for the cases
$\Omega_{\hbox{\scriptsize tot}} = 0.85, 0.9, 0.95$ and
$\Omega_{\hbox{\scriptsize tot}} = 1.0$ and compared
with the WMAP\protect\cite{Kogut_et_al_2003} data.
The model fits are determined by the minima of
$\chi_{\hbox{\scriptsize eff}}^2$
shown in Figures \ref{Fig:Omega_tot_85_2d} to \ref{Fig:Omega_tot_1_2d}.
}
\end{figure}
\begin{figure}[htb]
\begin{center}
\begin{minipage}{10cm}
\includegraphics[width=10cm]{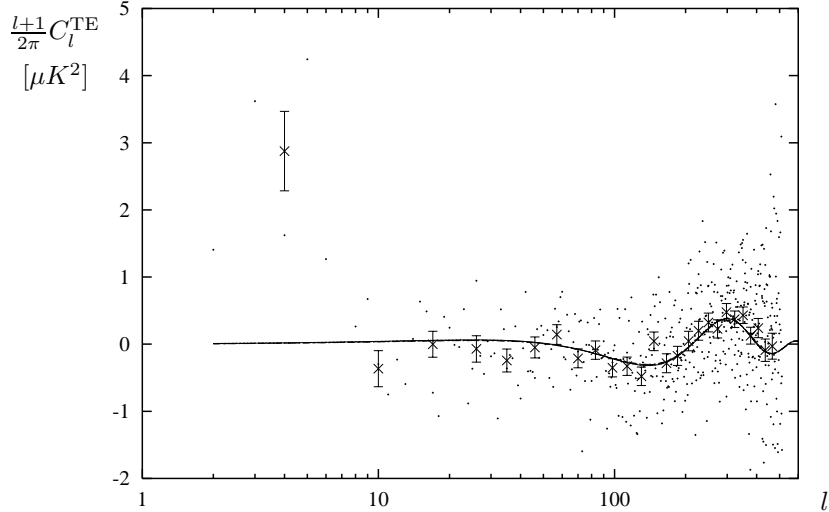}
\put(-5,4){$l$}
\put(-312,182){$\frac{l+1}{2\pi} C_l^{\hbox{\scriptsize{TE}}}$}
\put(-307,165){$[\mu K^2]$}
\end{minipage}
\vspace*{-10pt}
\end{center}
\caption{\label{Fig:aps_best_w_const_log_pol}
The same cross power spectra $(l+1) C_l^{\hbox{\scriptsize{TE}}}/(2\pi)$
as in Figure \ref{Fig:aps_best_w_const_lin_pol} are presented in a
logarithmic $l$-scale.
The large rise of the measured power for $l\lesssim 10$ is an indication
for a very early reionization.
}
\end{figure}

Before returning to the high optical depth $\tau$,
we would like to show in Figure \ref{Fig:aps_best_w_const_lin_pol}
that our models describe the TE-correlation $C_l^{\hbox{\scriptsize{TE}}}$
of the polarization with the temperature variations for $l \gtrsim 6$
very well.
(Note that there are no free parameters, since all parameters have already
been fixed by the TT power spectrum.)
Here the data obtained by WMAP \cite{Kogut_et_al_2003} are shown
in comparison with the four models discussed above
ignoring the reionization.
One observes that the oscillations are very well described by all four
models, and that the models are again degenerated.
In order to emphasize the oscillations,
we have chosen a scale in Figure \ref{Fig:aps_best_w_const_lin_pol}
which does not show some data points below $l\lesssim 5$.
In Figure \ref{Fig:aps_best_w_const_log_pol} we show the same
data as in Figure \ref{Fig:aps_best_w_const_lin_pol} using
a logarithmic scale.
One observes that the models do not describe the high observed values
below $l\lesssim 6$.
To describe these high values, a large optical depth $\tau$ is required.

\begin{figure}[htb]
\begin{center}
\begin{minipage}{10cm}
\includegraphics[width=10.0cm]{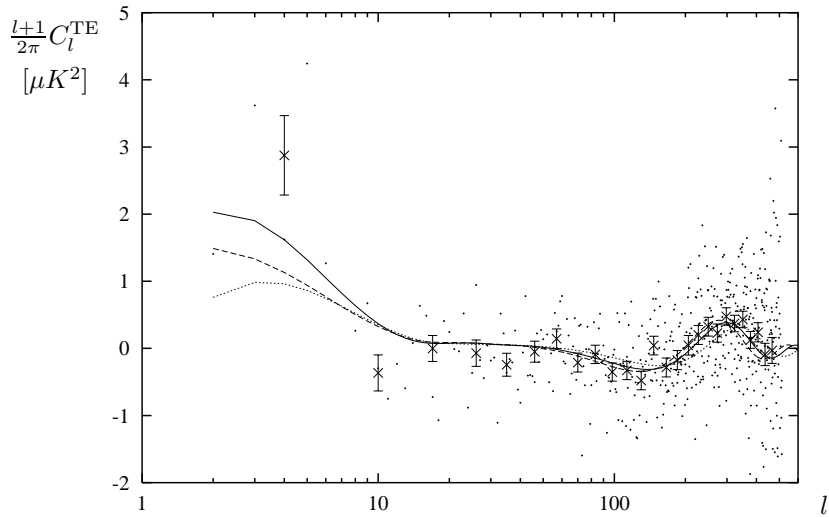}
\put(-5,4){$l$}
\put(-312,182){$\frac{l+1}{2\pi} C_l^{\hbox{\scriptsize{TE}}}$}
\put(-307,165){$[\mu K^2]$}
\end{minipage}
\vspace*{-10pt}
\end{center}
\caption{\label{Fig:aps_best_w_const_log_pol_influence}
The cross power spectra $(l+1) C_l^{\hbox{\scriptsize{TE}}}/(2\pi)$
are shown for three models with $\tau=0.15$,
$\Omega_{\hbox{\scriptsize b}} = 0.05$ and
$\Omega_{\hbox{\scriptsize cdm}} = 0.27$.
The solid curve belongs to a flat model with
$\Omega_{\hbox{\scriptsize tot}} = 1.0$,
$\Omega_\phi = 0.68$ and $w_\phi=-1.0$,
the dashed curve to a model which differs from the preceding one
only in the increased $w_\phi=-0.5$,
and the dotted curve belongs to a model where $\Omega_\phi$ is reduced
to $\Omega_\phi=0.58$, such that the model has
$\Omega_{\hbox{\scriptsize tot}} = 0.9$.
}
\end{figure}

Figure \ref{Fig:aps_best_w_const_log_pol_influence}
shows three models with a large optical depth $\tau=0.15$ and with
$\Omega_{\hbox{\scriptsize b}} = 0.05$ and
$\Omega_{\hbox{\scriptsize cdm}} = 0.27$.
The model belonging to the solid curve is a flat model,
i.\,e.\ having $\Omega_\phi = 0.68$,
with a cosmological constant $w_\phi=-1.0$.
This model shows the required increase for very small values of $l$.
Replacing the cosmological constant by a quintessence component
with $w_\phi=-0.5$ reduces the power as shown by the dashed curve
in Figure \ref{Fig:aps_best_w_const_log_pol_influence}.
Thus quintessence models require an even higher value of $\tau$
in order to obtain the same high level of TE-correlation
for low values of $l$.
The same behavior is observed if one changes instead of the
equation of state $w_\phi$ the curvature,
i.\,e.\ $\Omega_{\hbox{\scriptsize tot}}$.
The dotted curve belongs to a model which differs from the one
corresponding to the solid curve by a reduced $\Omega_\phi = 0.58$,
i.\,e.\ a model with $\Omega_{\hbox{\scriptsize tot}}=0.9$.
This curve reveals suppressed power at low values of $l$
in comparison to the flat case.
Models with negative curvature $\Omega_{\hbox{\scriptsize tot}}<1$
require also $w_\phi > -1$ in order to obtain for the first acoustic
peak a position around $l\simeq 220$,
as shown in Figure \ref{Fig:Peak_Postion}.
However, as we have just seen, both ingredients,
the increased $w_\phi$ as well as the negative curvature,
reduce the level of TE-correlation at low values of $l$.
To describe the observed TE-correlation, extremely high values of
$\tau$ are required,
and thus the reionization damping is in these cases too strong.
A Universe with negative curvature would then only be possible
if the multipoles at low values of $l$ are exceptional
outliers due to the cosmic variance.

The question whether quintessence models with negative curvature
are admissible or not thus depends on the optical depth $\tau$.
However, there is now some controversy whether a value of the order of
$\tau=0.17$ is compatible with physical models of structure formation.
In Refs.\ \cite{Haiman_Holder_2003,Cen_2003} it is shown that simple
models of reionization are inconsistent with both $\tau=0.17$
and the detection of the Gunn-Peterson trough in high-redshift quasars
\cite{Becker_et_al_2001,Fan_et_al_2003}.
In Ref.\ \cite{Chiu_Fan_Ostriker_2003} a degeneracy between
the optical depth $\tau$ and the spectral index $n_S$ is used
to argue for a lower value of $\tau\simeq 0.11$.
In this case a more complicated model involving a time-dependent
effective UV-efficiency is shown to be consistent
\cite{Chiu_Fan_Ostriker_2003} with the WMAP data.
In Ref.\ \cite{Ciardi_Ferrara_White_2003} N-body simulations
are presented which match the high value of $\tau$.
However, these simulations predict a mass-averaged neutral fraction
of $\sim 1 \%$ at $z\simeq 13$ and not at $z\simeq 6$ as required
by the high-redshift quasar spectra.
As a plausible solution of this problem, an epoch of recombination
following the first reionization is suggested
\cite{Ciardi_Ferrara_White_2003}
in order to have enough neutral matter at $z\simeq 6$.
Whether the current observed value of the optical depth withstands future data
remains to be seen.

\section{Summary}

In this paper, we have studied quintessence models with a constant
(effective) equation of state.
Special attention has been paid to the question
whether such quintessence models allow negative spatial curvature
with respect to the WMAP data.
The geometrical degeneracy for this model class is emphasized.
Furthermore, it is shown that a very high optical depth $\tau$
to the surface of last scattering would rule out models
with negative curvature.
If an optical depth around $\tau\simeq 0.17$ will be established,
then the Universe can be at most marginally negatively curved.
A flat Universe would then be established not by the position
of the acoustic peaks, which are equally well described by
quintessence models with negative curvature as we have shown,
but by the unexpected early reionization.
If on the other hand, the optical depth would be established as
negligibly small,
say around $\tau\simeq 0.05$, then there are quintessence models with
negative curvature which can describe the peak structure due to
the geometrical degeneracy in the three-dimensional
$(\Omega_{\hbox{\scriptsize tot}}, \Omega_\phi, w_\phi)$-space.


\section*{Acknowledgements}

We thank Antony Lewis and Antony Challinor for making their CAMB code
publicitly available.

\bibliographystyle{apsrev}
\bibliography{../../bib_astro}

\end{document}